\documentclass[aps,pra,graphicx,10pt]{revtex4-1}
\usepackage{amssymb,amsbsy,times,fancyhdr,color}
\usepackage{amsmath}
\usepackage{mathrsfs,amsfonts}
\usepackage{latexsym,afterpage}
\usepackage{graphicx,subfigure,multirow}
\usepackage{bbding}
\usepackage{color}
\usepackage{amsmath}
\newcommand\Rm{\mbox{Rm}} % Magnetic Reynolds number
\newcommand{\dy}{\partial}
\newcommand\ddy[2]{\frac{\dy#1}{\dy#2}}

\newcommand{\ex}{\mathrm{e}}

\newcommand{\grad}{\nabla}

\newcommand{\sech}{\mathrm{sech}}

\newcommand{\Bb}{\boldsymbol{B}}

\newcommand{\eb}{\boldsymbol{e}}

\newcommand{\jb}{\boldsymbol{j}}

\newcommand{\ub}{\boldsymbol{u}}

\definecolor{dark-green}{rgb}{0,0.5,0} % leeds

\begin{document}

% Use the \preprint command to place your local institutional report number 
% on the title page in preprint mode.
% Multiple \preprint commands are allowed.
%\preprint{}

\title{Vortex disruption by magnetohydrodynamic feedback} %Title of paper

% repeat the \author .. \affiliation  etc. as needed
% \email, \thanks, \homepage, \altaffiliation all apply to the current author.
% Explanatory text should go in the []'s, 
% actual e-mail address or url should go in the {}'s for \email and \homepage.
% Please use the appropriate macro for the type of information
% \affiliation command applies to all authors since the last \affiliation command. 
% The \affiliation command should follow the other information.

\author{J. Mak}
\email[]{julian.c.l.mak@googlemail.com}
\altaffiliation{School of Mathematics, The University of Edinburgh, James
Clerk Maxwell Building, The King's Buildings, Edinburgh, EH9 3FD, UK}

\author{S. D. Griffiths}

\author{D. W. Hughes}

%\email[]{Your e-mail address}
%\homepage[]{Your web page}
%\thanks{}
%\altaffiliation{}

\affiliation{Department of Applied Mathematics, University of Leeds, Leeds, LS2
9JT, UK}

% Collaboration name, if desired (requires use of superscriptaddress option in \documentclass). 
% \noaffiliation is required (may also be used with the \author command).
%\collaboration{}
%\noaffiliation
%\date{\today}

\begin{abstract}

In an electrically conducting fluid, vortices stretch out a weak, large-scale
magnetic field to form strong current sheets on their edges. Associated with
these current sheets are magnetic stresses, which are subsequently released
through reconnection, leading to vortex disruption, and possibly even
destruction. This disruption phenomenon is investigated here in the context of
two-dimensional, homogeneous, incompressible magnetohydrodynamics. We derive a
simple order of magnitude estimate for the magnetic stresses --- and thus the
degree of disruption --- that depends on the strength of the background magnetic
field (measured by the parameter $M$, a ratio between the Alfv\'en speed and a
typical flow speed) and on the magnetic diffusivity (measured by the magnetic
Reynolds number $\mbox{Rm}$). The resulting estimate suggests that significant
disruption occurs when $M^{2}\mbox{Rm} = O(1)$. To test our prediction, we
analyse direct numerical simulations of vortices generated by the breakup of
unstable shear flows with an initially weak background magnetic field. Using the
Okubo--Weiss vortex coherence criterion, we introduce a vortex disruption
measure, and show that it is consistent with our predicted scaling, for vortices
generated by instabilities of both a shear layer and a jet.

\end{abstract}

% insert suggested PACS numbers in braces on next line

\maketitle %\maketitle must follow title, authors, abstract and \pacs

% Body of paper goes here. Use proper sectioning commands. 
% References should be done using the \cite, \ref, and \label commands

%=====================================================================================
%=====================================================================================

%====================================================================

\section{Introduction}\label{sec:intro}

The interaction of vortices with a magnetic field is a fundamental process in
astrophysical magnetohydrodynamics (MHD). Such vortices could be generated, for
example, by convection \cite{Favier-et-al14, Guervilly-et-al14} or by the
breakup of unstable shear flows \cite{MashayekPeltier11, MashayekPeltier12a,
MashayekPeltier12b, Mak-et-al15}. In the absence of magnetic fields, vortices
can be coherent, long-lived structures, particularly in two-dimensional or
quasi-two-dimensional systems \cite[e.g.,][]{HoHuerre84}. However, in the
presence of a background magnetic field, various studies have shown how vortices
can be disrupted or completely destroyed \cite{Malagoli-et-al96, Frank-et-al96,
Jones-et-al97, Biskamp-et-al98, Keppens-et-al99, BatyKeppens02, Baty-et-al03,
BatyKeppens06, Palotti-et-al08}. Here we show explicitly how this disruption
depends on both the field strength and on the magnetic Reynolds number $\Rm$.

Astrophysical fluid flows are invariably characterised by extremely high values
of $\Rm$. Perhaps the most important consequence of this is that weak
large-scale fields can be stretched by the flow to generate strong small-scale
fields, with the amplification being some power of $\Rm$
\cite{VainshteinRosner91}. Once the small-scale fields are dynamically
significant then the resulting evolution is essentially magnetohydrodynamic,
rather than hydrodynamic, leading to dramatically different characteristics,
despite the large-scale magnetic field being very weak. Such behaviour has been
identified in the suppression of turbulent transport \cite{CattaneoVainshtein91,
KulsrudAnderson92, Tao-et-al93, GruzinovDiamond94, CattaneoHughes96,
Keating-et-al08, Kondic-et-al16}, in the suppression of jets in $\beta$-plane
turbulence \cite{Tobias-et-al07}, and in the inhibition of large-scale vortex
formation in rapidly rotating convection \cite{Guervilly-et-al15}.

The vortex disruption hypothesis investigated here depends on just such high
$\Rm$ dynamics. Given that many astrophysical flows are rotating and stratified,
such that the vortices are essentially two-dimensional, it is natural to
investigate vortex disruption in the context of two-dimensional MHD. In order to
quantify when a weak large-scale field can become dynamically significant, we
first construct a scaling argument for a quite general setting with a single
vortex. We follow the ideas of Weiss \cite{Weiss66}, first to estimate the
amplification of the large-scale field due to stretching on the edge of the
vortex, and then to estimate the resulting Lorentz force, which can disrupt the
radial force balance of the vortex. This leads to an explicit prediction for
vortex disruption in terms of the strength of the large-scale field and $\Rm$.

To investigate vortex disruption in more realistic settings it is necessary to
perform direct numerical simulations of the governing MHD equations for freely
evolving flows. The vortices may be imposed via the initial conditions or,
alternatively, may emerge from the development of an instability of a basic
state. Here we adopt the latter approach by considering the instability to
two-dimensional perturbations of two representative incompressible planar shear
flows -- a shear layer and a jet -- with an aligned magnetic field. This
typically leads to the generation of a periodic array of vortices, each of which
may be susceptible to disruption by the magnetic field. Any other potential
disruption mechanisms that require stratification (e.g., convective instability
in the vortex core due to overturning density surfaces \cite{MashayekPeltier12a,
MashayekPeltier12b}) or a third spatial dimension (e.g., elliptic instability
\cite{Kerswell02,Rieutord04} or the magnetoelliptic instability
\cite{LebovitzZweibel04,MizerskiBajer09}) are ruled out in our numerical
simulations, thereby isolating the magnetic field as the sole agent of vortex
disruption. 

Since magnetic dissipation is crucial in determining the eventual strength of
the small-scale fields, and hence the vortex disruption mechanism, it is
important to consider how this is implemented numerically. In order to have
well-defined values of $\Rm$, we carry out simulations with explicit Ohmic
dissipation, represented by a Laplacian, with resolution to the dissipative
scale. This is in contrast to most other studies of vortex disruption in MHD, in
which dissipation is performed numerically at the grid scale, with no explicit
diffusion operator \cite{Malagoli-et-al96, Frank-et-al96, Jones-et-al97,
Keppens-et-al99, BatyKeppens02, Baty-et-al03, BatyKeppens06}.

The plan of the article is as follows. Section~\ref{sec:theory} contains the
scaling argument for the disruption of a single vortex. The mathematical and
numerical formulation of the shear flow problem is given in
Section~\ref{sec:formulation}. Section~\ref{sec:disrupt} contains the numerical
simulations of vortex disruption; a measure of disruption based on the
Okubo--Weiss criterion \cite{Okubo70, Weiss91} is introduced and is used to test
the prediction of Section~\ref{sec:theory}. In Section~\ref{sec:dynamical} we
look at the implications for the large-scale dynamics of the shear flow
instability, with particular focus on the mean flow. We conclude in
Section~\ref{sec:conclusion}, and briefly discuss possible implications of
vortex disruption for the transition to turbulence and mixing.

%==============================================================================
%==============================================================================

\section{Theoretical estimate of vortex disruption}\label{sec:theory}

Following Weiss \cite{Weiss66}, we consider an idealised configuration in which
a single vortex evolves kinematically in an initially uniform magnetic field of
strength $B_0$. On an advective time scale, the fluid motion stretches the weak
background field on the edges of the vortex until reconnection of field lines
occurs; subsequently, the magnetic field lines within the vortex reconnect and
are expelled to the edges of the vortex, a phenomenon known as flux expulsion.
Weiss was particularly interested in determining the scalings of the
flux-expelled state, as were Moffatt \& Kamkar \cite{MoffattKamkar83} and, more
recently, Gilbert \emph{et al.} \cite{Gilbert-et-al16}. Here, however, we are
interested in the peak field at the point of reconnection at the edges of the
vortex (Weiss's $B_1$). Since the curved field lines have associated with them
magnetic stresses directed towards the vortex centre, then if $B_1$ is
sufficiently strong, the induced stresses will be significant, and we might
expect vortex disruption. We are then able to provide an estimate for the
dependence of disruption on $B_{0}$ and the magnetic diffusivity $\eta$, using
an essentially kinematic argument; put another way, the theory here estimates
when the kinematic approach breaks down and there must be significant dynamical
feedback. This approach is similar in spirit to that of Galloway and co-workers
\cite{Galloway-et-al78, GallowayMoore79}, who considered the dynamical feedback
of flux ropes formed in magnetoconvection.

The magnetic field is governed by the induction equation
\begin{equation}
	\frac{\dy\Bb}{\dy t}=\grad\times(\ub\times\Bb)+\eta\grad^{2}\Bb.
\end{equation}
We suppose that the vortex has characteristic length $L_v$ and velocity $U_v$.
The initial large-scale field $B_0$ is amplified to a stronger small-scale field
of strength $b$ and with a small characteristic scale $\ell$. Flux conservation
across the vortex implies that
\begin{equation}
\label{ch2:mag_amp1}
	b \ell = B_0 L_v.
\end{equation}
Field amplification is arrested when line stretching and magnetic diffusion are
in balance, i.e.\ when $|\Bb \cdot \nabla \ub | \sim \eta |\nabla^2 \Bb |$, or
$U_v B_0 /L_v \sim \eta b / \ell^2$. Combining this expression with
\eqref{ch2:mag_amp1} gives
\begin{equation}\label{ch2:mag_amp2}
	b\sim\left(\frac{U_v L_v}{\eta}\right)^{1/3}B_{0}.
\end{equation}
At this point, the magnetic stresses resulting from the Lorentz force have magnitude
\begin{equation}
	\left| \jb \times \Bb \right| = \left|\frac{1}{\mu_{0}}(\grad\times\Bb)\times\Bb\right|
	\sim \frac{1}{\mu_0}\frac{b^2}{\ell} \sim
	\frac{U_v}{\mu_0} \frac{B_0^2}{\eta},
\label{eq:LF1}
\end{equation}
where $\mu_0$ is the permeability of free space, and we have used the facts that
$b \gg B_0$ and $\ell \ll L_v$.

In the hydrodynamic case, the centripetal force on the vortex is $\rho U_v^2 /
L_v$, where $\rho$ is the (constant) density. This suggests that magnetic
stresses will be important when
\begin{equation}
	\left| \frac{1}{\mu_0} (\grad \times \Bb) \times \Bb \right|
	\sim \frac{\rho U^2_v}{L_v}.
\label{eq:LF2}
\end{equation}
The disruptive regime is characterised by equating expressions~\eqref{eq:LF1}
and \eqref{eq:LF2}, leading to
\begin{equation}\label{ch2:disruption}
	\frac{U_v}{\mu_0} \frac{B_0^2}{\eta} \sim \frac{\rho U^2_v}{L_v}
	\qquad \Rightarrow \qquad
	\frac{B_0^2/ \left( \mu_0 \rho \right)}{\eta} \sim \tilde {\Omega}_v,
\end{equation}
where $\tilde \Omega_v \sim U_v/L_v$ is the dimensional vorticity.

Throughout this derivation we have assumed that $B_0$ is sufficiently weak that
the Lorentz force does not come into play before the field amplification is
arrested diffusively. Furthermore, we are assuming that disruption of the
vortex, if it occurs, does so on a much faster time scale than that for reaching
the flux-expelled state.%, namely $\Rm^{1/3} L_v/U_v$.

Expression \eqref{ch2:disruption} is the dimensional estimate for vortex
disruption in terms of the characteristic scales of the vortex. Typically,
however, there are other velocity and length scales, $U_0$ and $L_0$, that are
used to characterise the flow. Retaining $B_0$ as the characteristic field
strength, the relevant non-dimensional parameters are
\begin{equation}\label{eq:mRm}
	M = \frac{B_0/\sqrt{\mu_0 \rho}}{U_0}	
	\qquad \Rm = \frac{U_0 L_0}{\eta},
\end{equation}
where $M$ is the ratio of the Alfv\'en speed to the characteristic flow speed,
and is a measure of field strength. Expression \eqref{ch2:disruption} may thus
be written as 
\begin{equation}
M^2 \Rm \sim \Omega_v,
\end{equation}
where
\begin{equation}
\Omega_v = \frac{U_v/L_v}{U_0/L_0}
\end{equation}
is the non-dimensional magnitude of the vortex. In many settings, $L_v$ and
$U_v$ will be comparable with $L_0$ and $U_0$, in which case the non-dimensional
estimate for vortex disruption becomes
\begin{equation}\label{eq:disruption_nondim}
M^2 \Rm \sim 1.
\end{equation}

%==============================================================================
%==============================================================================

\section{Mathematical and numerical formulation}\label{sec:formulation}

To examine vortex disruption in detail, we consider the evolution of unstable
shear flows of the form $\ub = U(y) \eb_{x}$, in the presence of a uniform
background magnetic field $\Bb=B_{0}\eb_{x}$, in two-dimensional incompressible
MHD. Since both $\ub$ and $\Bb$ are divergence-free, they may be expressed in
terms of a streamfunction and magnetic potential, defined by
\begin{equation}
	\ub =(u,v,0) = \grad\times(\psi\eb_{z}) , \quad \quad
	\Bb/\sqrt{\mu_{0}\rho}=\grad\times(A\eb_{z}).
\end{equation}
The $z$-components of the vorticity and current are then given by
$\omega=-\grad^2 \psi$ and $j = -\grad^2 A$ respectively. On scaling velocity
with a representative flow speed $U_0$, length with a characteristic scale
$L_0$, time with $L_0/U_0$, and magnetic field with $B_0$, the non-dimensional
governing equations become
\begin{subequations}\label{ch3:MHD-stream}
\begin{align}
	\frac{\dy\omega}{\dy{}t}-J(\psi,\omega)-M^{2}J(A,\grad^{2}A)
	&=\frac{1}{\mbox{Re}}\grad^{2}\omega, \label{ch3:MHD-stream-a}\\
	\frac{\dy A}{\dy{}t}-J(\psi,A)
	&=\frac{1}{\mbox{Rm}}\grad^{2}A, \label{ch3:MHD-stream-b}\\
	-\grad^{2}\psi&=\omega \label{ch3:MHD-stream-c},
\end{align}
\end{subequations}
where
\begin{equation}
	J(f,g)=\frac{\dy f}{\dy x}\frac{\dy g}{\dy y}
	-\frac{\dy f}{\dy y}\frac{\dy g}{\dy x}
\end{equation}
is the Jacobian operator. The non-dimensional parameters are $M$ and $\Rm$, as
defined in \eqref{eq:mRm}, and the Reynolds number
\begin{equation}
	\mbox{Re}=\frac{U_{0}L_{0}}{\nu},
\end{equation}
where $\nu$ is the kinematic viscosity. In non-dimensional form, the two flow
profiles we shall consider are $U(y)=\tanh(y)$ and $U(y)=\sech^{2}(y)$, which we
shall refer to as the shear layer and the jet respectively. 

We further decompose the variables in terms of a basic state and a perturbation,
i.e.,
\begin{equation}
	\psi=\Psi(y)+\tilde{\psi},\qquad A=y+\tilde{A},\qquad 
	\omega=-U'(y)+\tilde{\omega},
\end{equation}
where a prime denotes differentiation with respect to $y$. The system of
equations \eqref{ch3:MHD-stream} then takes the equivalent formulation (after
dropping the tildes on the perturbation terms)
\begin{subequations}\label{ch3:MHD-stream-split}
\begin{align}
	\frac{\dy\omega}{\dy{}t}+U\frac{\dy\omega}{\dy{}x}
	-U''\frac{\dy\psi}{\dy{}x}-J(\psi,\omega)
	-M^{2}\left[-\frac{\dy\grad^{2}A}{\dy{}x}+J(A,\grad^{2}A)\right]
	&=\frac{1}{\mbox{Re}}\grad^{2}\omega-\frac{1}{\mbox{Re}}U''',
	\label{ch3:MHD-stream-split-a}\\
	\frac{\dy{}A}{\dy{}t}+U\frac{\dy{}A}{\dy{}x}-\frac{\dy\psi}{\dy{}x}
	-J(\psi,A)
	&=\frac{1}{\mbox{Rm}}\grad^{2}A, \label{ch3:MHD-stream-split-b}\\
	-\grad^{2}\psi&=\omega. \label{ch3:MHD-stream-split-c}
\end{align}
\end{subequations}
We adopt a domain that is periodic in $x$ and bounded by impermeable, perfectly
conducting and stress-free walls at $y=\pm L_{y}$, leading to the boundary
conditions
\begin{equation}
	\psi=0,\qquad A=0,\qquad \omega=0 \qquad\textnormal{on}\qquad y=\pm L_y.
\end{equation}
The total energy (the sum of the kinetic energy $E_k$ and the magnetic energy
$E_m$) decays as
\begin{equation}\label{eq:energy-equation}
	\frac{\mathrm{d}}{\mathrm{d}t} (E_k+E_m) 
	%\iint (E_{k}+E_{m})\ \mathrm{d}x\mathrm{d}y
	=-\frac{1}{\mbox{Re}}\iint\omega^{2}\ \mathrm{d}x\mathrm{d}y
	-\frac{1}{\mbox{Rm}}\iint j^{2}\ \mathrm{d}x\mathrm{d}y,
\end{equation}
where $E_k$ and $E_m$ are the domain integrals of $|\ub|^2/2$ and
$M^2|\Bb|^{2}/2$, respectively. 

The channel length is chosen to be some integer multiple of the most unstable
wavelength from linear theory, i.e., $L_x = 2n\pi/\alpha_c$, where $\alpha_c$ is
the most unstable wavenumber of the associated profiles ($\alpha_c = 0.44$ for
the shear layer and $\alpha_c = 0.90$ for the jet \cite{DrazinReid-Stability}).
We take $n=1$ for the shear layer and $n=2$ for the jet, giving domains of
roughly equal size. To trigger the instability, we initialise with a
perturbation for which the primary instability at wavenumber $\alpha_c$ has
fixed amplitude and phase, with other permitted wavenumbers $\alpha_i$ having
smaller amplitude and random phase. Specifically, we take
\begin{equation}\label{eq:perturb}
	(\omega,A) = \left[10^{-3}\cos(\alpha_c x)+
	10^{-5}\sum_{i\neq n}^{\lfloor N_{x}/3\rfloor}
	\gamma_{i}\cos(\alpha_i x - \sigma_i L_x)\right]\ex^{-y^{2}},
\end{equation}
where $\gamma_i$ and $\sigma_i$ are randomly generated numbers in $[-1,1]$, and
$\lfloor \cdot \rfloor$ is the floor function. There is a well-defined linear
phase of instability, during which the most unstable eigenfunction naturally
emerges. Simulations were run up to $t=150$, allowing for an extended nonlinear
phase and the possibility of vortex disruption; with this choice we find that
taking $L_y = 10$ ensures that finite boundary effects remain negligible.

We solve the system of equations \eqref{ch3:MHD-stream-split} by a
Fourier--Chebyshev pseudo-spectral method, using the standard Fourier
collocation points in $x$ and Gauss--Lobatto points in $y$, and employing the
appropriate Fast Fourier Transforms. A semi-implicit treatment in time is
employed, treating the dissipation terms implicitly and the nonlinear terms
explicitly. Time-stepping is performed by a third-order accurate, variable
time-step, Adams--Bashforth/Backward-Difference scheme, with the step size set
by the maximum time step allowed for a fixed CFL number (here taken to be $0.2$,
which is near marginal for numerical stability). The equations are solved in
spectral space using a fast Helmholtz solver \cite{Thual-thesis}, and all runs
are dealiased using Orszag's $2/3$-rule \cite{Orszag71b} (see
\cite{Canuto-et-al-Spectral} or \cite{Peyret-Spectral} for further details about
the numerical methods employed).

Our simulations are run-down experiments for the evolution of instabilities on a
decaying background state. To alleviate diffusive effects before the
perturbations reach finite amplitude, we remove the diffusion of the basic state
(the $U'''$ term in \eqref{ch3:MHD-stream-split-a}) until the perturbation is
sufficiently large (here measured by the energy possessed by the $k_x \neq 0$
Fourier modes). Even so, we found it necessary to take $\mbox{Re} \gtrsim 500$
in order to produce runs that are not too diffusive and that are qualitatively
similar to the runs at higher $\mbox{Re}$. Since the regime
estimate~\eqref{eq:disruption_nondim} naturally suggests a dependence on $M$ and
$\mbox{Rm}$, we fix $\mbox{Re} = 500$ and vary the other two parameters in the
bulk of this work. The required spatial resolution depends on $\mbox{Rm}$: the
number of $x$-gridpoints $N_x$ and $y$-gridpoints $N_y$ were taken to be $N_x
\times N_y = 512 \times 1024$ for $\mbox{Rm} = 1000$, $384 \times 768$ for
$\mbox{Rm} = 750$, and $256 \times 512$ otherwise.

%==============================================================================
%==============================================================================

\section{Vortex disruption}\label{sec:disrupt}

In this section, we describe the results of direct numerical simulations of
freely evolving vortices generated by shear instabilities with a background
magnetic field. To measure the disruption of the vortices, we follow Okubo
\cite{Okubo70} and Weiss \cite{Weiss91} in considering the quantity $W(x,y,t)$,
defined by
\begin{equation}\label{eq:okubo-weiss}
  W = \left(\ddy{u}{x} - \ddy{v}{y}\right)^2 +
    \left(\ddy{v}{x} + \ddy{u}{y}\right)^2 -
    \left(\ddy{v}{x} - \ddy{u}{y}\right)^2 .
\end{equation}
The bracketed terms are respectively the normal and shear components of the rate
of strain, and the vorticity; $W$ thus effectively measures the relative
dominance of the strain over the vorticity. A vortex is defined as a region in
which $W$ is sufficiently negative. For example, a popular approach is to
calculate the standard deviation $\sigma$ of $W$ and to classify vortical
regions by $W < -0.2 \sigma$. Though by no means the only way to identify a
vortex \citep[e.g.,][]{Haller05, Haller-et-al16}, it is one of the simpler
measures that have been employed previously in a geophysical setting
\cite[e.g.,][]{IsernFotanet-et-al06, Waugh-et-al06}.

Here we are interested in measuring vortex disruption relative to the purely
hydrodynamic evolution. We thus introduce a vortex disruption parameter
$\Delta(t)$, defined by
\begin{equation}\label{eq:ok-measure}
  \Delta = 1 - \frac{\int_A W\, \mathrm{d}x\, \mathrm{d}y}{\int_A W_{\tiny \mbox{HD}}\, \mathrm{d}x\, \mathrm{d}y} ,
\end{equation}
where $W_{\tiny \mbox{HD}}$ is the value of $W$ for the hydrodynamic case. The
area $A$ is some portion of physical space where there is deemed to be a vortex,
as discussed in more detail below. When $\Delta = 0$, the vortex is not
disrupted, whereas $\Delta = 1$ implies total disruption.

%---------------------------------------------------------------

\subsection{Hyperbolic-tangent shear layer}\label{sec:tanh-disrupt}

We first consider the disruption of vortices arising from the instability of the
shear layer $U(y)=\tanh(y)$. Linear instabilities of this profile are well
documented for both the hydrodynamic \cite{DrazinReid-Stability} and
MHD cases, where linear stability is guaranteed when $M \geq 1$
\cite{Vladimirov-et-al96, HughesTobias01}. The nonlinear MHD evolution has been
studied by numerous authors \citep[e.g.,][]{Frank-et-al96, Malagoli-et-al96,
Jones-et-al97, Keppens-et-al99, Baty-et-al03, Palotti-et-al08}. In order to
provide a good coverage of $(M, \Rm)$ space, a total of fifty simulations with
parameter values given in Table~\ref{table:tanh-param-values} were performed,
at five values of $\mbox{Rm}$ and ten values of $M$, where the values of $M$ are
chosen to lie well below the stabilising value of $M$ ($M \approx 0.8$ for
$\alpha=0.44$). The initial magnetic to kinetic energy ratio is given by
$2M^{2}L_{y}/(\int U(y)^2 \ \mathrm{d}y) \approx (10/9) M^2$; this ratio is less
than $1/90$ for the values of $M$ considered here.

\begin{table}[tbp]
\centering
\begin{tabular}{cccc}
	\toprule
	parameter & values & \hspace*{10mm} & marker/colour in Fig.~\ref{fig:tanh_okubo_weiss_regime}\\
	\colrule
	$M$ & 0.01--0.1, 0.01 spacing & & $+\ \circ\ \ast\ \times\ \Box\ \Diamond\ \bigtriangleup\ \bigtriangledown\ \text{\FiveStarOpen}\ \text{\DavidStar}$, in ascending order\\
	$\mbox{Rm}$ & 50, 250, 500, 750, 1000 & & black, red, green, blue, magenta\\
		\botrule
\end{tabular}
\caption{Parameter values employed for the shear layer simulations.}
\label{table:tanh-param-values}
\end{table}

\begin{figure}[tbp]
\begin{center}
	\includegraphics[width=0.8\textwidth]{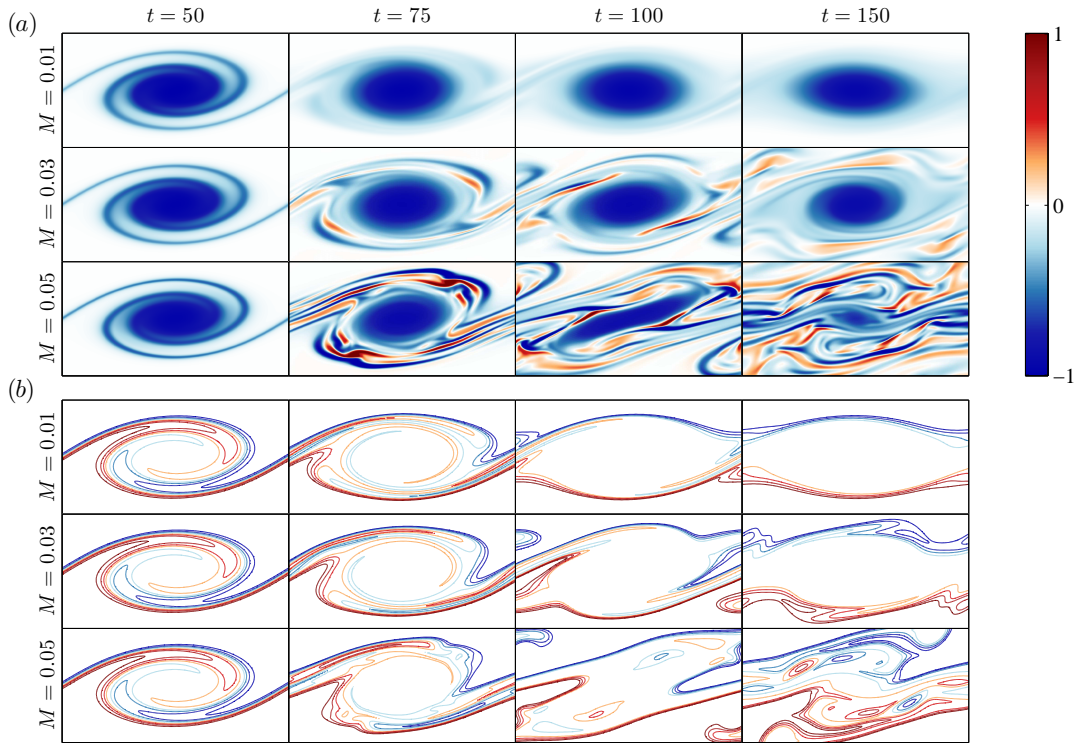} 
	\caption{Snapshots of ($a$) vorticity and ($b$) magnetic field lines for the
	shear layer at different field strengths (for $\mbox{Rm}=\mbox{Re}=500$),
	shown for the central half of the channel ($-L_y/2 \le y \le L_y/2$).}
	\label{fig:vorticity_time_series_tanh_MHD}
\end{center}
\end{figure}

We consider a domain supporting a single wavelength of the optimum linear
instability mode. Fig.~\ref{fig:vorticity_time_series_tanh_MHD} shows snapshots
of the vorticity and magnetic field lines from three control runs at $\Rm =
500$. These display the representative behaviours for three dynamical regimes:
undisturbed ($M = 0.01$), mildly disrupted ($M = 0.03$) and severely disrupted
($M = 0.05$). For $M=0.01$, the vorticity is of one sign, and the shear layer
rolls up into a vortex. The magnetic stresses are clearly not strong enough to
alter the macro-dynamics in any significant way. The vortex evolution is
essentially hydrodynamic \cite{HoHuerre84}, accompanied by magnetic flux
expulsion from the vortex. For $M=0.03$, we observe the formation of regions of
positive vorticity. The magnetic field is no longer confined to kinematic
boundary layers, and the resulting stresses are strong enough to modify the
resulting evolution to a certain extent. That said, the vortex is only mildly
disrupted and maintains its integrity; there is only a slight decrease of vortex
size by the end of the simulation at $t=150$. For $M=0.05$, the evolution is
radically different to the other two cases, with a significant disruption of the
vortex and an unconfined magnetic field. By the end of the simulation, only
small remnants of the parent vortex persist; vorticity filaments and a complex
magnetic field are now the dominant features in the domain.

\begin{figure}[tbp]
\begin{center}
	\includegraphics[width=0.8\textwidth]{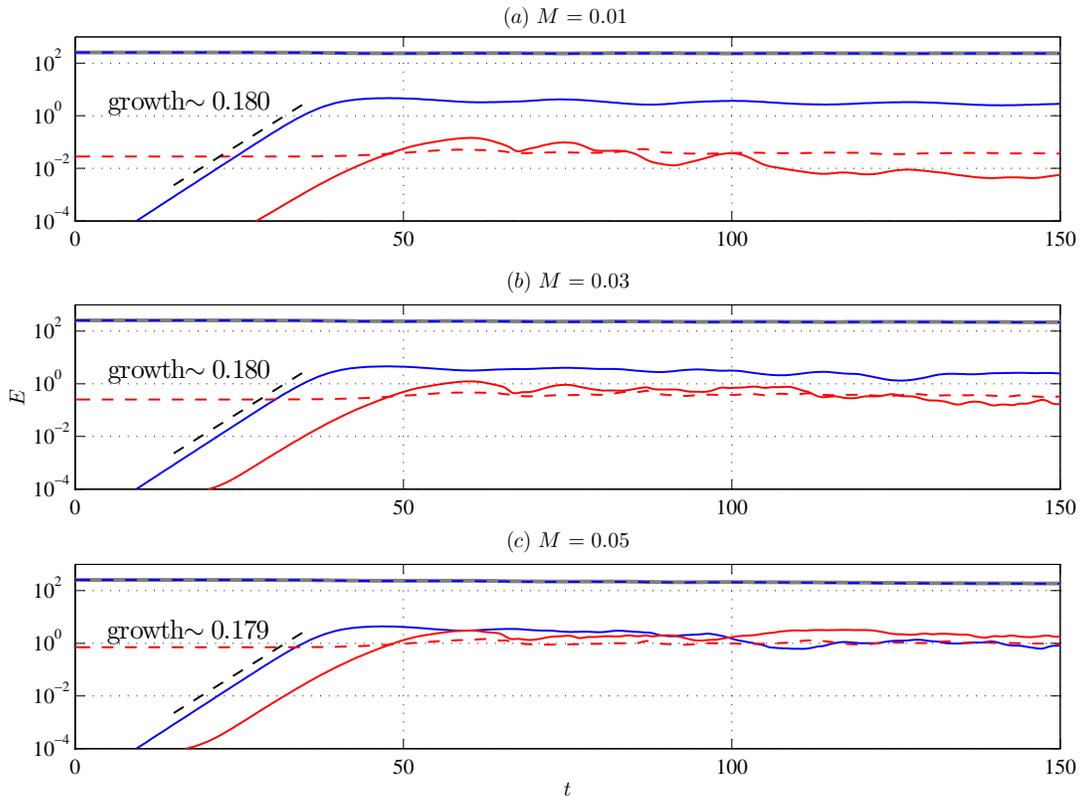}
	\caption{Time series of the energies (blue $=$ kinetic; red $=$ magnetic; solid $=$
	perturbation state; dashed $=$ mean state; grey solid line $=$ total energy)
	for the shear layer at different field strengths (for
	$\mbox{Rm}=\mbox{Re}=500$). The curve for mean kinetic energy largely lies on
	top of the curve for the total energy.}
	\label{fig:energy_total_tanh_MHD}
\end{center}
\end{figure}

Vortex disruption also has a signature in the time evolution of the kinetic and
magnetic energies. Shown in Fig.~\ref{fig:energy_total_tanh_MHD} are time series
for the three control runs of the mean kinetic energy $\overline{E}_k$ and mean
magnetic energy $\overline{E}_m$ (defined as the energy content in the $k_{x}=0$
Fourier mode), along with the perturbation energies $E'_k$ and $E'_m$ (defined
as the energy content in the remaining Fourier modes). The evolution is similar
up to $t \approx 60$ (cf.\ Fig.~\ref{fig:vorticity_time_series_tanh_MHD}), at
which time the field amplification is close to being arrested by diffusion; the
scalings (\ref{ch2:mag_amp1}) and (\ref{ch2:mag_amp2}) then apply for the
small-scale field, implying $E'_m \sim b^2 l L_v$ and $\overline{E}_m \sim B_0^2
L_0^2$, so that $E'_m / \overline{E}_m \sim {\rm Rm}^{1/3} \approx 8$ here,
consistent with Fig.~\ref{fig:energy_total_tanh_MHD}. However, for $t \gtrsim
80$, vortex disruption (if it occurs) changes the evolution of the energy. For
the undisrupted case ($M=0.01$), the evolution becomes one of complete flux
expulsion (see Fig.\ref{fig:vorticity_time_series_tanh_MHD}), with $E'_m$
decreasing to less than $\overline{E}_m$. (This is different to the well-known
theory of \cite{Weiss66}, in which $E'_m \sim {\rm Rm}^{1/2} \overline{E}_m$ in
the flux-expelled state, but that kinematic single-vortex theory may not apply
to this dynamic regime with a periodic array of vortices and remote boundaries.)
For the strongly disrupted case ($M=0.05$), we enter a different regime, with
$E'_m$ staying close to $\overline{E}_m$ throughout the evolution. This period
with persistent small spatial scales leads to higher dissipation: whereas the
total dissipation is small and comparable with that of the hydrodynamic case for
$M=0.01$ and $0.03$, it is about three times higher when $M=0.05$. Further, even
though $\overline{E}_m \ll \overline{E}_k$ throughout the evolution for all
three cases (i.e., weak large-scale magnetic field), $E'_m$ becomes comparable
with $E'_k$ when there is vortex disruption, reflecting the dynamical importance
of the small-scale magnetic field.

\begin{figure}[tbp]
\begin{center}
	\includegraphics[width=0.7\textwidth]{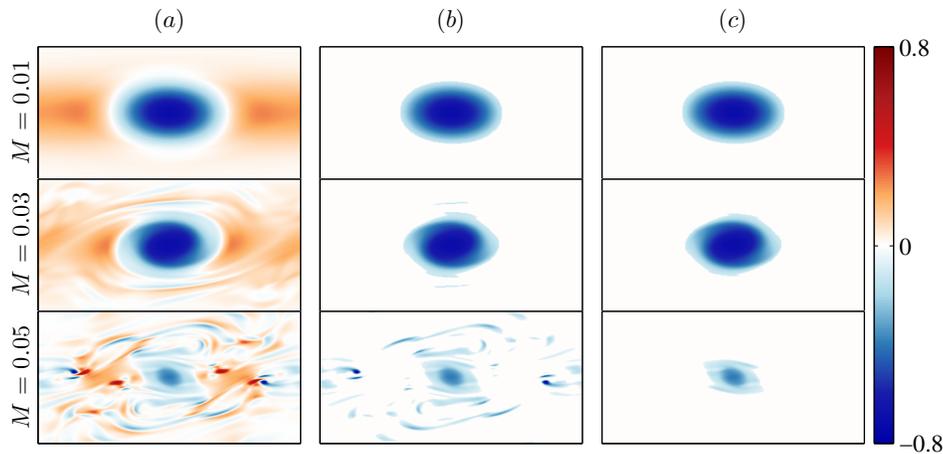}
	\caption{Okubo--Weiss field for the shear layer corresponding to the $t=150$
	cases shown in Figure~\ref{fig:vorticity_time_series_tanh_MHD}. ($a$) The full
	$W$ field given in equation \eqref{eq:okubo-weiss}; ($b$) the field keeping
	only $W < -0.2\sigma$; ($c$) a further filtered field keeping only the largest
	connected region originating from the centre of the parent vortex.}
	\label{fig:okubo_weiss_tanh}
\end{center}
\end{figure}

To calculate the vortex disruption parameter $\Delta$, defined by
\eqref{eq:ok-measure}, we first need to evaluate $W(x,y,t)$. This is shown in
Fig.~\ref{fig:okubo_weiss_tanh}($a$) for the three control runs at $t=150$,
highlighting those regions where either strain or vorticity dominates. Adopting
the convention of identifying vortical regions as those where $W < -0.2 \sigma$
results in the plots of Fig.~\ref{fig:okubo_weiss_tanh}($b$). For $M=0.01$ and
$M=0.03$, this leads to the identification of a well-defined coherent vortex.
For the severely disrupted case of $M = 0.05$ on the other hand, disruption
results in a small parent vortex, as well as disconnected vortices and
filaments. Our interest is in the disruption to the parent vortex, and so to
employ the disruption measure $\Delta$, we need to ignore these resulting
by-products. To address this, a further filter is applied, which selects the
largest connected region originating from the centre of the parent vortex (which
in this case is well defined since the resulting instability has zero phase
speed, so the single vortex formed is stationary). The $W$ field is digitised,
with all points where $W<-0.2\sigma$ set to one, and all other points set to
zero. The MATLAB command \verb|bwlabel| is then applied to the resulting data
array to select the connected regions; the connected region originating from the
centre of the vortex is then chosen as the region $A$ for the calculation of
$\Delta$ (see Fig.~\ref{fig:okubo_weiss_tanh}($c$)).

\begin{figure}[tbhp]
\begin{center}
	\includegraphics[width=0.7\textwidth]{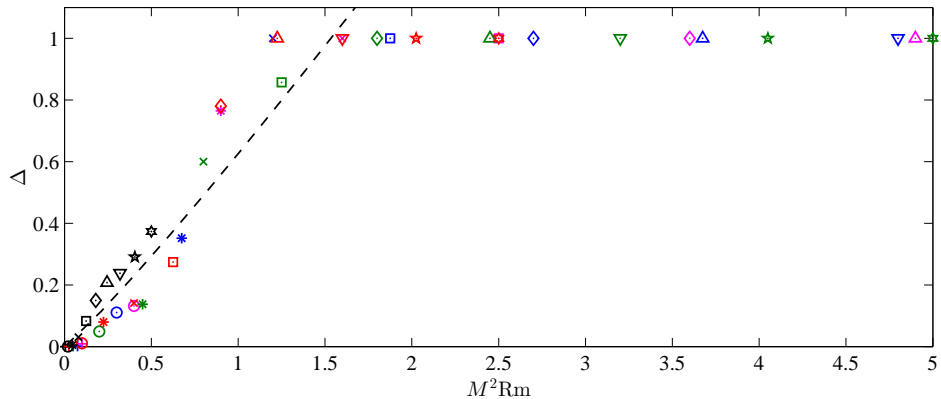}
	\caption{$\Delta$ vs $M^2 \Rm$ for the shear layer (colours: varying
	$\mbox{Rm}$; markers: varying $M$; see Table~\ref{table:tanh-param-values} for
	marker and colour values). The dotted line is obtained from a regression of
	the data. For display purposes, data beyond $M^2\mbox{Rm} = 5$ is omitted.}
	\label{fig:tanh_okubo_weiss_regime}
\end{center}
\end{figure}

The quantity $\Delta$ is computed at $t=150$ for all simulations detailed in
Table~\ref{table:tanh-param-values}. To test the vortex
prediction~\eqref{eq:disruption_nondim}, in
Fig.~\ref{fig:tanh_okubo_weiss_regime} we show $\Delta$ versus $M^2 \Rm$, from
which it can be seen that there is a reassuring collapse of the data. The most
important point to note is that for $M^2 \Rm \gtrsim 1.5$, the vortices are
deemed to have been completely destroyed. This is in agreement with
\eqref{eq:disruption_nondim}, which predicts disruption for $M^2 \Rm \sim 1$.
Further, for $M^2 \Rm \lesssim 1$ we would expect $\Delta$ to be monotonically
increasing with $M^2 \Rm$. Indeed, on performing a regression on the data in the
interval $0.1 \leq \Delta \leq 0.9$, we find that $\Delta \sim \left( M^2 \Rm
\right)^{1.10}$ (although a different definition of $\Delta$ could lead to a
different positive exponent).

Our focus here has been on the transition from an essentially kinematic regime
to a dynamic regime when $M^2 \sim {\rm Rm}^{-1} \ll 1$. If $M^2$ is increased
beyond ${\rm Rm}^{-1}$, then severe vortex disruption continues for a while, as
implied by Fig.~\ref{fig:tanh_okubo_weiss_regime}. However, if $M$ becomes
sufficiently large ($M \gtrsim 0.35$ here), then the magnetic field is
sufficiently strong that vortex formation is completely suppressed.

%---------------------------------------------------------------

\subsection{Bickley jet}\label{sec:sech-disrupt}

We now consider the disruption of vortices arising from the instability of the
Bickley jet, $U(y)=\sech^{2}(y)$. Linear instabilities of this profile in the
hydrodynamic setting are again well documented \cite{DrazinReid-Stability};
there are odd and even modes of instability, with the latter being the most
unstable at wavenumber $\alpha_c = 0.90$. In the MHD setting, linear stability
is guaranteed for this configuration when $M \geq 0.5$ \cite{HughesTobias01}.
This jet profile is known to break up into vortices if the initial field is not
so strong that it suppresses the primary hydrodynamic instability; depending on
the parameter values, the resulting vortices have been observed to suffer
disruption by the magnetic field \cite{Biskamp-et-al98, BatyKeppens02,
BatyKeppens06}.

The optimum wavenumber $\alpha_c = 0.90$ is roughly twice that of the shear
layer case. In order to employ the same resolution as the shear layer case, we
consider a channel length that is twice this optimum wavelength. Fifty
simulations with parameter values given in Table~\ref{table:sech-param-values}
were carried out, with the values of $M$ again chosen to lie well below the
stabilising value of $M$ ($M \approx 0.3$ for $\alpha=0.90$, implying lower
values of $M$ than for the shear layer). For the Bickley jet, the initial
magnetic to kinetic energy ratio is given by $2M^{2}L_{y}/(\int U(y)^2 \
\mathrm{d}y) \approx 15 M^2$; this ratio is less than $3/80$ for the values of
$M$ considered here.

\begin{table}[tbhp]
\centering
\begin{tabular}{cccc}
	\toprule
	parameter & values & \hspace*{10mm} & marker/colour in Fig.~\ref{fig:sech_okubo_weiss_regime}\\
	\colrule
	$M$ & 0.005--0.05, 0.005 spacing & & $+\ \circ\ \ast\ \times\ \Box\ \Diamond\ \bigtriangleup\ \bigtriangledown\ \text{\FiveStarOpen}\ \text{\DavidStar}$, in ascending order\\
	$\mbox{Rm}$ & 50, 250, 500, 750, 1000 & & black, red, green, blue, magenta\\
	\botrule
\end{tabular}
\caption{Parameter values employed for the jet simulations.}
\label{table:sech-param-values}
\end{table}

\begin{figure}[tbp]
\begin{center}
	\includegraphics[width=0.8\textwidth]{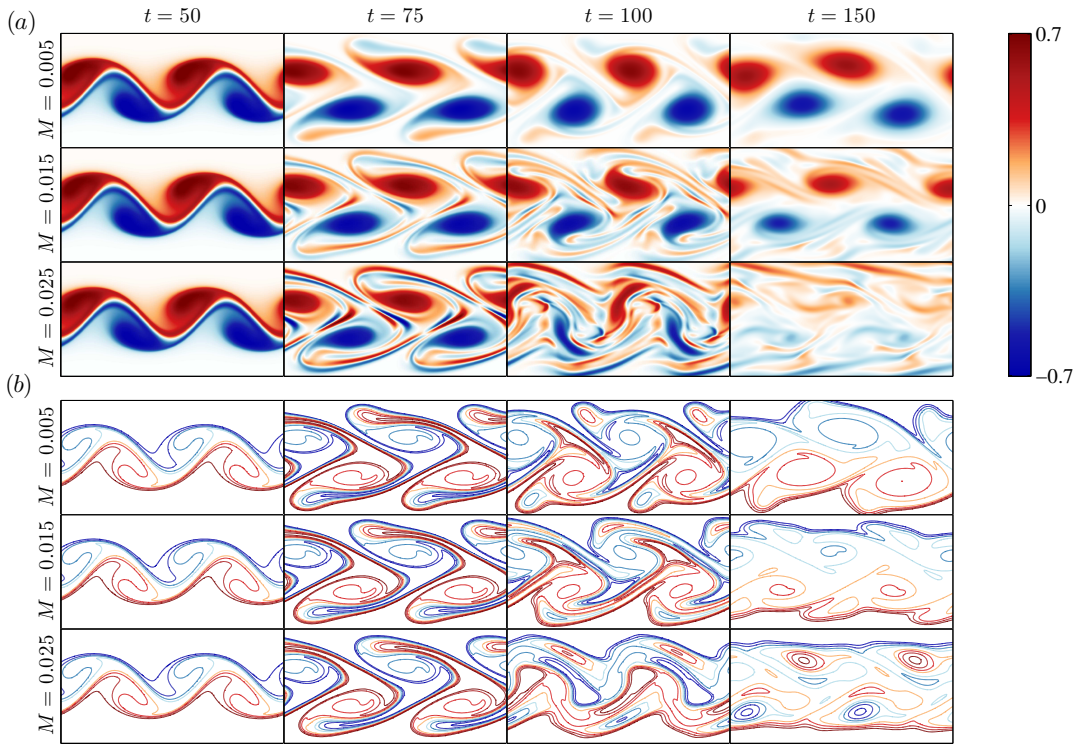}
	\caption{Snapshots of ($a$) vorticity and ($b$) magnetic field lines for the
	jet at different field strengths (for $\mbox{Rm}=\mbox{Re}=500$), shown for
	the central half of the channel ($-L_y/2 \le y \le L_y/2$).}
	\label{fig:vorticity_time_series_sech_MHD}
\end{center}
\end{figure}

Three control runs are again chosen, with $\mbox{Rm}=500$ and $M=0.005$, $0.015$
and $0.025$. Fig.~\ref{fig:vorticity_time_series_sech_MHD} shows snapshots of
the vorticity and magnetic field lines for the three representative cases. For
$M=0.005$, the evolution is essentially hydrodynamic, with a meandering of the
jet before it breaks into two pairs of vortices. MHD feedback is weak and there
are no visible disruptions to the vortices. For $M=0. 015$, the vortices at
$t=100$ are slightly distorted by the released magnetic stresses; disruption,
however, is not strong enough to destroy the vortices. For $M=0.025$, the
induced magnetic stresses are strong enough to distort the vortices
significantly; indeed, the vortex cores have almost disappeared by $t=150$.

\begin{figure}[tbp]
\begin{center}
	\includegraphics[width=0.8\textwidth]{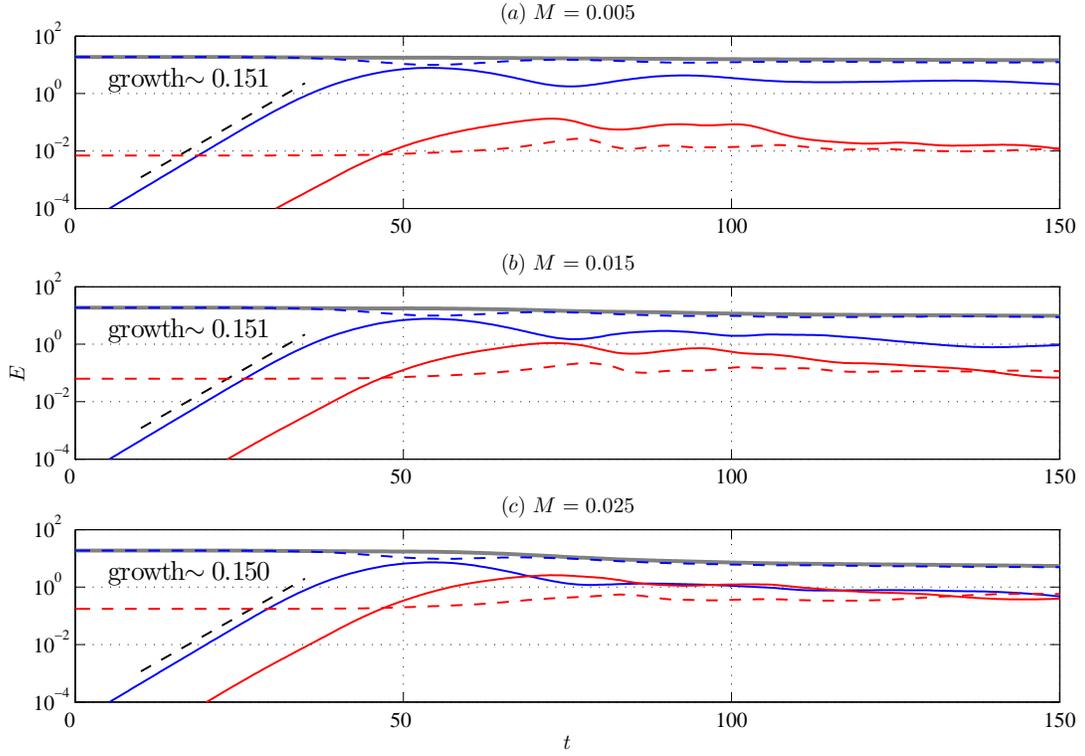}
	\caption{Time series of the energies (blue $=$ kinetic; red $=$ magnetic;
	solid $=$ perturbation state; dashed $=$ mean state; grey solid line $=$ total
	energy) for the jet at different field strengths (for
	$\mbox{Rm}=\mbox{Re}=500$).}
	\label{fig:energy_total_sech_MHD}
\end{center}
\end{figure}

Fig.~\ref{fig:energy_total_sech_MHD} shows the time series of the energies for
the three control runs. The relative sizes of $\overline{E}_k$, $E'_k$,
$\overline{E}_m$ and $E'_m$ follow the same patterns as for the shear layer.
However, in contrast to the shear layer, here the total dissipation is actually
lower for the severe vortex disruption case ($M=0.025$) than for the undisrupted
case ($M=0.005$). As shown in Fig.~\ref{fig:vorticity_time_series_sech_MHD},
although the small-scale structures in the vorticity and magnetic field are
initially amplified by severe disruption ($t=100$), which enhances the
dissipation, for longer times they are suppressed ($t=150$). 

\begin{figure}[tbp]
\begin{center}
	\includegraphics[width=0.7\textwidth]{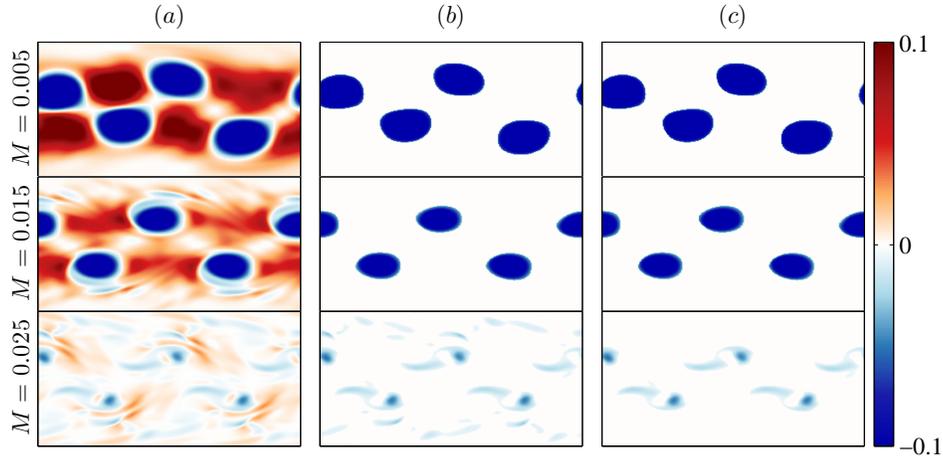}
	\caption{Okubo--Weiss field for the jet corresponding to the $t=150$ cases
	shown in Fig.~\ref{fig:vorticity_time_series_sech_MHD}. ($a$) The full $W$
	field given in equation \eqref{eq:okubo-weiss}; ($b$) the field keeping only
	$W < -0.2\sigma$; ($c$) a further filtered field keeping only the four largest
	connected components, accounting for periodicity. The colour scale is
	saturated to show the small values when $M=0.025$.}
	\label{fig:okubo_weiss_sech}
\end{center}
\end{figure}

\begin{figure}[tbp]
\begin{center}
	\includegraphics[width=0.8\textwidth]{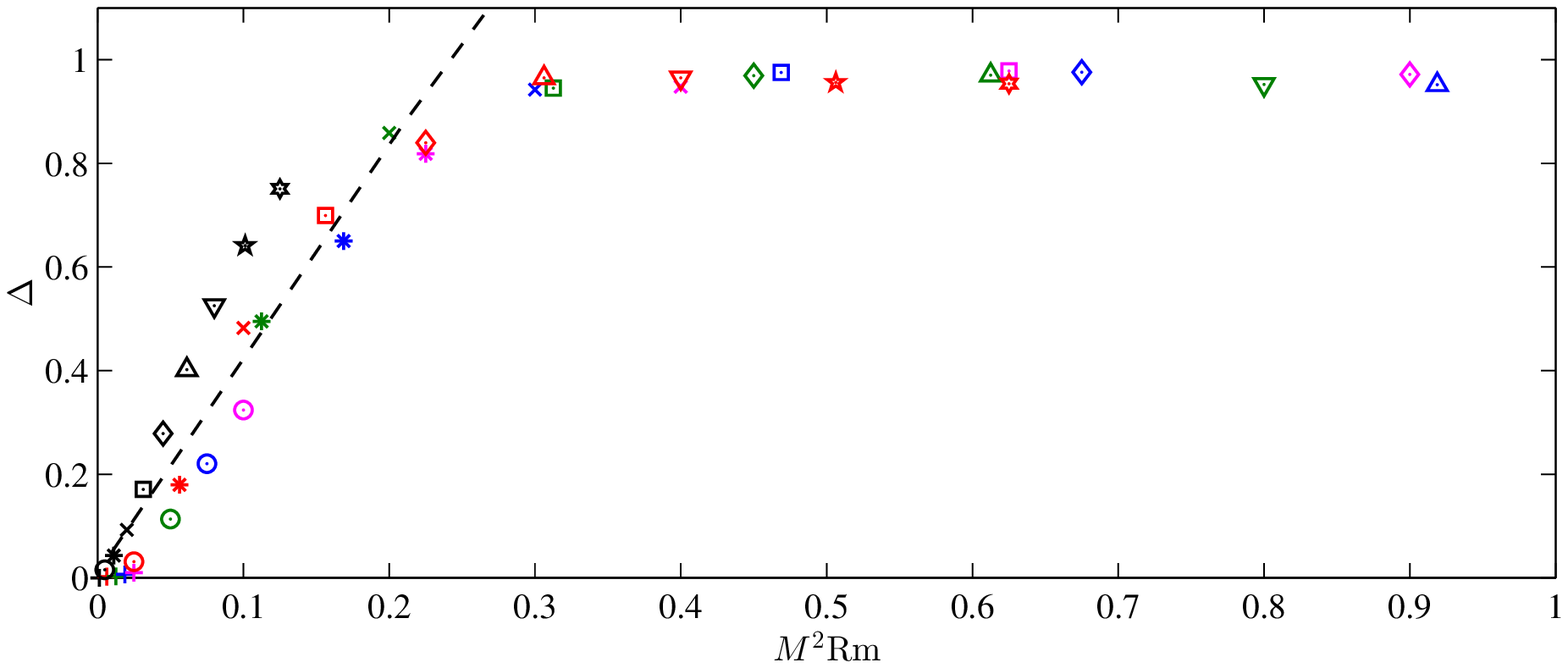}
	\caption{$\Delta$ vs $M^2 \Rm$ for the jet (colours: varying $\mbox{Rm}$;
	markers: varying $M$; see Table~\ref{table:sech-param-values} for marker and
	colour values). The dotted line is obtained from a regression of the
	data. For display purposes, data beyond $M^2\mbox{Rm} = 1$
	is omitted.}
	\label{fig:sech_okubo_weiss_regime}
\end{center}
\end{figure}

We calculate $\Delta$ using a similar procedure as for the shear layer. However,
since four vortices are generated by the instability in this configuration, to
calculate the area $A$ for the integral \eqref{eq:ok-measure} we now select the
four largest connected components of $W(x,y)$, accounting for periodicity. The
procedure is illustrated for the three control runs in
Fig.~\ref{fig:okubo_weiss_sech}, at $t=150$. When performed for all fifty
simulations given in Table~\ref{table:sech-param-values}, we obtain the plot of
$\Delta$ (at $t=150$) vs $M^2 \Rm$ given in
Fig.~\ref{fig:sech_okubo_weiss_regime}. It shows the same general
characteristics as for the shear layer: there is an approximately linear
increase of $\Delta$ with $M^2 \Rm$ ($\Delta \sim (M^2\mbox{Rm})^{0.94}$) up to
a critical value, given by $M^2 \Rm \approx 0.3$, above which there is complete
vortex disruption ($\Delta \approx 1$). Note that for both the shear layer and
the jet, the critical value of $M^2 \Rm$ for vortex disruption is of order
unity, although the precise value varies from case to case.

%==============================================================================

\section{Dynamical phenomena}\label{sec:dynamical}

We believe that the vortex disruption hypothesis, leading to the
estimate~\eqref{eq:disruption_nondim}, is quite general. However, in any given
setting, there could be interesting dynamical implications beyond those of the
disruption itself. Here, we spell out some of the specific implications for the
nonlinear evolution of shear flow instabilities.
%---------------------------------------------------------------

\subsection{Mean flow changes}\label{sec:mean}

A quantity of considerable importance in the instability of shear flows is the
mean flow, defined here to be
\begin{equation}
	\overline{u}(y,t)=\frac{1}{L_{x}}\int_{0}^{L_{x}}u(x,y,t)\ \mathrm{d}x,
\end{equation}
where $L_x$ is the channel length. In combination with the magnetic field, the
mean flow determines the strength of the initial linear instability, and its
evolution with time shows how the nonlinear dynamics act to mix momentum in the
cross-stream direction. 
  
\begin{figure}[tbhp]
\begin{center}
	\includegraphics[width=0.8\textwidth]{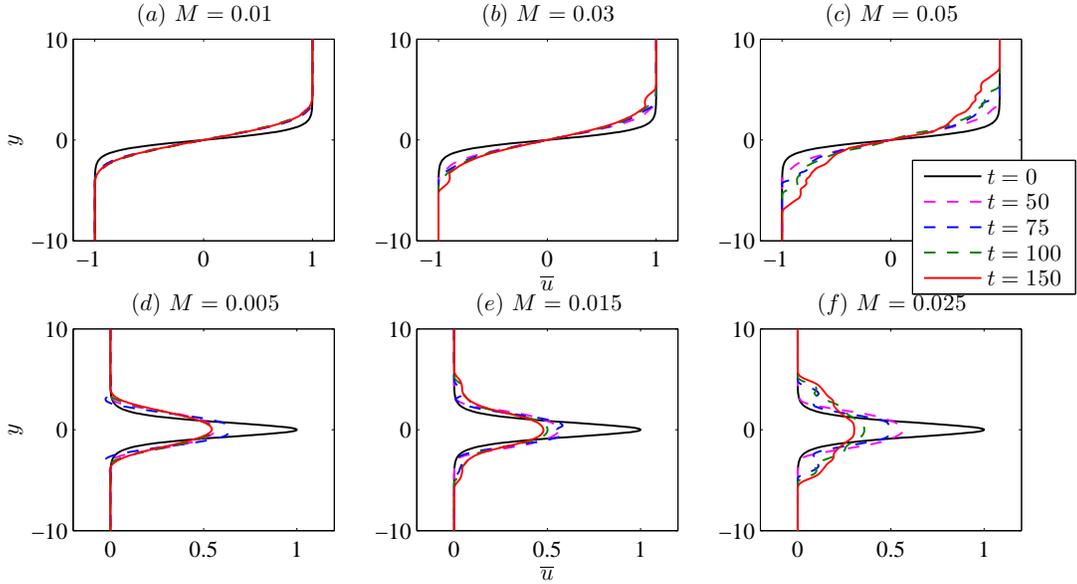}
	\caption{Snapshots of $\overline{u}$ for the shear layer ($a$, $b$, $c$) and
	jet ($d$, $e$, $f$) at different field strengths (for
	$\mbox{Rm}=\mbox{Re}=500$). Some of the curves lie on top of each other and
	are indistinguishable.}
	\label{fig:mean_flow}
\end{center}
\end{figure}
  
Snapshots of $\overline{u}$ for the shear layer and the jet are shown in
Fig.~\ref{fig:mean_flow}, for each of the three control runs. For the
undisturbed cases (essentially hydrodynamic) in Fig.~\ref{fig:mean_flow}($a,d$),
the shear is reduced around the centre of the channel as the instability reaches
finite amplitude, but the mean flow remains largely unchanged thereafter. For
the mildly disrupted cases in Fig.~\ref{fig:mean_flow}($b,e$), the behaviour is
similar (i.e., essentially hydrodynamic) near the centre of the channel ($|y|
\lesssim 2$), but beyond that there is an additional broadening of the mean flow
when vortex disruption sets in, for $t \gtrsim 100$. However, for the severely
disrupted cases in Fig.~\ref{fig:mean_flow}($c,f$), the mean-flow evolution is
substantially different, with mixing of momentum over a wider region (about
twice as wide as for the hydrodynamic case), leaving smaller shears near the
centre of the channel. This is consistent with previous studies
\cite[e.g.,][]{Palotti-et-al08}. 

It is possible to quantify the influence of vortex disruption on the evolving
mean flow, by measuring the width of the mean-flow changes relative to those of
the hydrodynamic case~\cite{Mak-thesis}. This analysis reveals how the width of
the changes to the mean-flow increases with $M^2 {\rm Rm}$, with substantial
widening when $M^2 {\rm Rm} \sim 1$, i.e., within the vortex disruption regime. 

%---------------------------------------------------------------

\subsection{Secondary hydrodynamic instabilities}\label{sec:secondary}

It is well known that vortices generated during the finite-amplitude stage of
shear instabilities can be unstable to a range of secondary hydrodynamic
instabilities. Our numerical simulations were designed to suppress such
secondary hydrodynamic instabilities, so that only magnetic disruption of the
vortices could occur. However, within our two-dimensional system, there is the
possibility of the subharmonic pairing instability \cite{Kelly67, Jimenez87,
Jimenez88}, which requires two or more vortices in the along-stream direction.
For our shear layer simulations, this was completely suppressed, since only a
single vortex was generated within the domain. For our jet simulations, where
two vortices were generated in the along-stream direction, early signatures of
the subharmonic pairing instability may be seen in
Fig.~\ref{fig:vorticity_time_series_sech_MHD} for the undisturbed case with
$M=0.005$, and to a lesser extent in the mildly disrupted case with $M=0.015$.
However, in the strongly disrupted case with $M=0.025$, the magnetic disruption
occurs on a faster time scale than the pairing instability, and the disrupted
vortices show no signs of pairing. An interesting alternative scenario is when
magnetic disruption does not act on the primary vortices, but in which repeated
subharmonic pairings eventually lead to a large vortex that does suffer magnetic
disruption. Although there is evidence of this occurring in studies of longer
channels \cite{Baty-et-al03}, we did not find this in sample simulations where
the domain was extended to allow for eight wavelengths of the primary
instability.  

For different flow configurations, other hydrodynamic secondary instabilities
could occur. With a third spatial dimension, there is the possibility of either
hyperbolic instabilities on the thin braids between vortices or ellipitical
instabilities on the vortex cores, as discussed in \cite{CaulfieldPeltier00},
for example. With density stratification, there is also the possiblity of
convective type instabilities associated with density overturning in the vortex
cores, which compete with various other
modes~\cite{MashayekPeltier12a,MashayekPeltier12b}. However, such secondary
instabilities are all excluded here through our choice of a two-dimensional
system of constant density. In three-dimensional or stratified flows, whether or
not such secondary instabilities would act before magnetic disruption of the
parent vortices is an open question. 

%==============================================================================

\section{Conclusion and discussions}\label{sec:conclusion}

Of great astrophysical significance is the idea that a very weak large-scale
magnetic field, i.e.\ a field with energy much smaller than the kinetic energy
of the flow in question, can still have dynamically significant consequences.
Small-scale, typically turbulent, motions distort the large-scale field to
generate small-scale magnetic fields, amplifying the field strength by some
positive power of the magnetic Reynolds number $\Rm$. Since the defining
characteristic of astrophysical plasmas is that $\Rm \gg 1$, this implies that
weak large-scale magnetic fields cannot simply be ignored. This phenomenon has
been explored for the suppression of both turbulent magnetic diffusivity
\cite{VainshteinRosner91, CattaneoVainshtein91, Keating-et-al08, Kondic-et-al16}
and the turbulent $\alpha$-effect of mean field electrodynamics
\cite{KulsrudAnderson92, Tao-et-al93, GruzinovDiamond94, CattaneoHughes96}, the
inhibition of jet formation in $\beta$-plane turbulence \cite{Tobias-et-al07},
and the suppression of large-scale vortices in rapidly rotating convection
\cite{Guervilly-et-al15}. Here we have explored how, in another broad class of
problems --- the destruction of vortices by a weak large-scale field --- the
same general dynamical processes can occur. By adopting the arguments of
\cite{Weiss66}, we are able to estimate the magnitude of the induced magnetic
stresses arising from the stretching of magnetic field lines by the swirling
fluid motion. Vortices are disrupted when these magnetic stresses exceed the
centripetal forces for the vortex; this occurs when $M^2 \Rm \sim 1$, where
$M^2$ measures the energy of the large-scale field relative to the kinetic
energy. Thus when $\Rm \gg 1$, vortex disruption occurs for $M \ll 1$.

The estimate for vortex disruption, $M^2 \Rm \sim 1$, is widely applicable for
vortex dynamics, however the vortices may arise. Here we test the criterion in
detail by considering one of the most important means of vortex generation ---
the nonlinear development of shear instabilities. We focus on two-dimensional,
homogeneous, incompressible MHD, considering in detail two prototypical shear
flows (the hyperbolic tangent shear layer and the Bickley jet) with a weak
background magnetic field. The instability of both these flows leads to a
periodic array of vortices, which can be prone to magnetic disruption. We first
identify coherent vortices by the Okubo--Weiss procedure, which provides a
simple definition of vortices as those regions where vorticity dominates over
strain, and then construct a measure of disruption $\Delta$ by comparison with
the purely hydrodynamic evolution. By performing fifty simulations to cover a
range of $M$ ($0.01 \le M \le 0.1$ for the shear layer and $0.005 \le M \le
0.05$ for the jet) and $\Rm$ ($50 \le \Rm \le1000$) at $\mbox{Re}=500$, we are
able to investigate the dependence of $\Delta$ on both $M$ and $\Rm$. For both
shear flows, we find an approximate linear increase of $\Delta$ with $M^2 \Rm$
up to a critical value of $M^2 \Rm$ of order unity, followed by a sharp
transition to a regime in which $\Delta \approx 1$, denoting total disruption of
the vortices. These numerical results are in excellent agreement with the
theoretical estimate. Our theoretical ideas are also supported by inspection of
the energy time series, which show that disruption is characterised by an
ordering in which the energies of the perturbed magnetic field, the mean field,
and the perturbed flow are comparable (but all much less than the kinetic energy
of the unperturbed shear flows).

The vortex disruption estimate depends crucially on a balance between advection
and dissipation of small-scale field. Since dissipation is represented by a
Laplacian operator in the induction equation,
estimate~\eqref{eq:disruption_nondim} involves an explicit dependence on $\Rm$.
This dependence is captured by our numerical approach, which uses a Laplacian
diffusion operator with resolution down to the dissipation scale. Numerical
schemes with alternative (non-Laplacian) prescriptions for the dissipation would
presumably realise vortex disruption in a somewhat different way.

The notion of vortex disruption has some interesting astrophysical implications.
Since the derived estimate for disruption is $M^{2}\mbox{Rm} \sim 1$, and
$\mbox{Rm}$ is typically extremely large in astrophysical systems, vortex
disruption should be a robust dynamical feature. If vortex disruption does
occur, it is likely to lead to mixing of quantities such as angular momentum,
heat and passive scalars, with implications for the large-scale angular
velocity, temperature and chemical composition. Of particular note is that in
systems such as the solar tachocline, constrained by stable stratification, many
of the standard mixing scenarios do not occur. However, provided there are
vortices, we have shown that their interaction with a magnetic field can provide
an alternative route to mixing. The theoretical disruption
estimate~\eqref{eq:disruption_nondim} assumes that $\Rm \gg 1$ and implicitly
assumes that the flow is smooth on the small scales of the field; this equates
to an assumption that the magnetic Prandtl number
$\mbox{Pm}=\nu/\eta=\mbox{Rm}/\mbox{Re} \gtrsim O(1)$. Whereas this does indeed
hold in the interstellar medium, in stellar interiors $\mbox{Pm} \ll 1$.
Although one could envisage vortex disruption by a similar physical mechanism in
this regime, the line of argument leading to a disruption estimate would need to
be modified. Furthermore, the testing of any such hypothesis in the regime
$\mbox{Re} \gg \mbox{Rm} \gg 1$ is currently computationally unattainable.

%==============================================================================

\begin{acknowledgments}

This work was supported by the STFC Doctoral Training Grant ST/F006934/1.

% Put your acknowledgments here.

\end{acknowledgments}

%=====================================================================================
%=====================================================================================

\bibliography{refs}

\end{document}